\documentclass[sigconf]{acmart}

\usepackage{booktabs} 
\usepackage[disable]{todonotes}
\usepackage{color}
\usepackage[export]{adjustbox}

\usepackage{microtype}

\usepackage{geometry} 
\usepackage{graphicx} 

\usepackage{float} 

\usepackage{amsmath} 

\usepackage{multicol}
\usepackage{multirow}

\usepackage{paralist}

\usepackage{url}

\usepackage{comment}

\usepackage{rotating}

\usepackage[toc,page]{appendix}

\graphicspath{{Pictures/}} 

\usepackage{pgf} 
\usepackage{tikz} 
\usetikzlibrary{arrows,automata,shapes} 
\usetikzlibrary{calc}
\usetikzlibrary{positioning}

\usepackage{enumitem}

\usepackage{cleveref}

\setcopyright{none}

\begin{document}
\title{Secure Evaluation of Knowledge Graph Merging Gain}


\author{Leandro Eichenberger}
\affiliation{%
	\institution{Computer Science 5, RWTH Aachen}
	\city{Aachen} 
	\country{Germany} 
}
\email{leandro.eichenberger@rwth-aachen.de}

\author{Michael Cochez}\authornote{Most of the work on this paper has been done while working at Fraunhofer FIT}
\orcid{0000-0001-5726-4638}
\affiliation{%
	\institution{Computer Science, Vrije Universiteit Amsterdam}
	\city{Amsterdam} 
	\country{the Netherlands} \\
}
\email{m.cochez@vu.nl}

\author{Benjamin Heitmann}
\orcid{????????????/}
\affiliation{%
	\institution{Computer Science 5, RWTH Aachen}
\city{Aachen} 
\country{Germany} \\
	\institution{Fraunhofer FIT}
	\city{Aachen} 
	\country{Germany} \\
}
\email{benjamin.heitmann@cs.rwth-aachen.de}

\author{Stefan Decker}
\orcid{????????????/}
\affiliation{%
	\institution{Computer Science 5, RWTH Aachen}
	\city{Aachen} 
	\country{Germany} \\
	\institution{Fraunhofer FIT}
	\city{Aachen} 
	\country{Germany} \\
}
\email{stefan.decker@rwth-aachen.de}


\begin{abstract}

Finding out the differences and commonalities between the knowledge of two parties is an important task.
Such a comparison becomes necessary, when one party wants to determine how much it is worth to acquire the knowledge of the second party, or similarly when two parties try to determine, whether a collaboration could be beneficial.
When these two parties cannot trust each other (for example, due to them being competitors) performing such a comparison is challenging as neither of them would be willing to share any of their assets.
This paper addresses this problem for knowledge graphs, without a need for non-disclosure agreements nor a third party during the protocol. 

During the protocol, the intersection between the two knowledge graphs is determined in a privacy preserving fashion.
This is followed by the computation of various metrics, which give an indication of the potential gain from obtaining the other parties knowledge graph, while still keeping the actual knowledge graph contents secret.
The protocol makes use of blind signatures and (counting) Bloom filters to reduce the amount of leaked information.
Finally, the party who wants to obtain the other's knowledge graph can get a part of such in a way that neither party is able to know beforehand which parts of the graph are obtained (i.e., they cannot choose to only get or share the good parts). 
After inspection of the quality of this part, the Buyer can decide to proceed with the transaction.

The analysis of the protocol indicates that the developed protocol is secure against malicious participants.
Further experimental analysis shows that the resource consumption scales linear with the number of statements in the knowledge graph.
\end{abstract}

%
%



\maketitle

\todo[inline]{MC The CCSXML still has to be checked -- one removed to make alignment nice, other title suggestions welcome, keywords to be improved}
\todo[inline]{MC Clean up references.}
\todo[inline]{MC Add a citation to thesis in final version}
\todo[inline]{MC Check cross refs}
\todo[inline]{MC Check capitalization: Buyer, Seller, Bloom}

\linespread{0.9}

\section{Introduction}

\todo[inline]{1 page}

For many people and businesses most or at least a big part of their value lies in their knowledge.
Often this knowledge is represented in knowledge graphs.
This is true for a broad range of companies ranging from social media to medical.
For any business, it is crucial to be able to buy, sell, and trade.  
For these businesses, it is important to do so with their knowledge.
However, unlike any physical goods, knowledge, and in this case a knowledge graph, is copied and replicated very easily at no cost.
Therefore, a Seller does not want to share his knowledge graph to a potential Buyer before having a deal because he has no guarantee that the Buyer won't just keep a copy of the graph when the deal is canceled.
The Buyer, on the other hand, would like to integrate new information into his own knowledge graph.
He does, however, not want to blindly buy a knowledge graph without knowing how much information it contains beyond what he already knows.

This creates the need to privately compare two knowledge graphs, giving privacy guarantees to the Seller and enabling the Buyer to evaluate how much knowledge he can gain and therefore how much he should pay for the knowledge graph in question.
Informally, the protocol needs to guarantee Seller privacy, meaning that the Buyer learns no more information about the Seller's knowledge graph than the two agreed upon.
For the Buyer, the guarantees are that the Seller learns very little about the Buyer's knowledge graph and cannot influence the outcome.
Since the quality of data can be measured in many, case dependent ways, it is necessary that the Buyer can obtain a part of the Seller's knowledge graph.
However, it must be ensured that neither party can select what part it is to avoid a biased selection.

The main contribution of this article is the protocol for privacy preserving Knowlegde Graph merging gain measurement.
Its most profound benefits are:

\begin{itemize}[leftmargin=*]
	\item The ability to compute explicitly what is in the intersection of the Seller's and Buyer's graph and metrics which give an indication of the amount of new information in the merger of the graph, with a small, and analyzed, amount of exposed information shared between the parties (see \cref{sec:approach,sec:Leaks}).
	\item The possibility to share a part of the knowledge graph from the Seller to the Buyer such that neither party can make a biased selection (see \cref{sec:approach:quality})
	\item The ability to verify that the Seller has followed the protocol as intended (see \cref{sec:approach:verfication}).
	\item A practically linear use of resources: time, space, and communication (see \cref{sec:Evaluation} and specifically \cref{chart:runtime-all,chart:memory_all,fig:communicationBtoS,fig:communicationStoB})
\end{itemize}
We provide the implementation of the protocol and evaluation
on github \url{https://github.com/blindedwebconf/submission1688}.

\section{Comparing Knowledge Graphs Without Sharing Them}
\todo[inline]{1 page}


The goal of this paper is to find a way, to compare two knowledge graphs of different parties, without having to show them to one another.
The goal of this comparison is, for one party (`the Buyer') to figure out if and how much it could benefit from obtaining the knowledge graph of the other party (`the Seller').
To achieve this, it is important to find the intersection of the two knowledge graphs.
Besides, we want to be able to share specific statistics on the Seller's knowledge graph, as well as compute metrics which indicate the potential information gain.
These so called entropy metrics are computed by the Buyer on its own graph and on a proxy for the merged graphs, ensuring the Buyer does only learn a small amount of information about the Seller's graph. 
The Buyer wants to get new high quality knowledge from the acquired graph.
However, what knowledge and quality means is use case and user specific.
Hence, there is a need to provide a part of the Seller's knowledge graph to the buyer.
As neither of the parties can be trusted to make the selection of the part of the graph to be shared, this should happen such that the selection is essentially random (i.e., beyond control of either party).

As both parties want to keep as much information private as possible, it is also crucial to find measure for the amount of information leaked during the process.
These metrics will show the amount of information leaked for each individual step of the protocol.
As various parts of the protocol do not depend on each other, the user has the possibility to decide, which information he is willing to compute at the cost of a certain information leak.
This protocol is meant to be usable in real world scenarios, therefore it needs to be sufficiently efficient and usable.
This includes scaling to larger knowledge graphs, even if it is reasonable to assume that two major companies trying to work out a big deal would provide substantial processing power and be are willing to spend more time, if this ensures privacy and precise information on the possible gain.
Trusted third parties are undesirable for this protocol because they can be hard to find and eliminating them also eliminates the risk of collusion.


One possible area of use for this protocol is medical companies.
In~\cite{mccusker2017finding} a knowledge graph is used to find new potential drugs against metastatic cutaneous melanoma, which is an aggressive skin cancer.
The knowledge graph contains drug-protein, protein-protein, and gene-disease interactions.
In this case they found 25 new candidates for new drugs that fulfill certain criteria when searching the knowledge graph.
In this case, it is easy to see why someone would be interested in buying such a knowledge graph.
Also, this makes it quite clear why privacy is paramount.
The knowledge graph used in this case was built of data open to the public, but medical companies invest huge sums into research and a new cancer drug is worth a lot.

Another area where this protocol is beneficial is the trade of user data, where this protocol can be used to determine whether the Seller and Buyer 
have an overlapping user base and whether they the types of information they have on users is different.

A different, more general case is any two companies wanting to explore, if a cooperation between them would make sense.
In this case they can run the protocol in both directions.
This way they can figure out, whether the respectively other party has different knowledge supplementing their own, which makes a cooperation beneficial.

As the ideas used in this protocol still work in case one knowledge graph is empty, it is also possible for a Buyer, which is new to a certain field, to run this protocol with a potential Seller.
This allows the Buyer, to learn whether buying the Seller's knowledge graph could be a good entry into this field or not.

\subsection{Requirements}

Before going over the steps and ideas of this protocol we will first specify its requirements.
There are two aspects to this.
First, the protocol needs to be usable in a real-world scenario and second, it has cryptographic parts to it which need to give certain guarantees. 

\noindent\textbf{Information gain:} This protocol is meant to help a Buyer to decide if it is beneficial to him to buy a certain knowledge graph.
This means, the Buyer must be able to estimate the amount of information he could gain from a merger.
This creates the need to measure the difference and the similarities between the knowledge graphs.

\noindent\textbf{Data quality:} An important part of making the decision whether to buy a knowledge graph or not is testing its quality.
While the protocol does not in itself need to test the data quality, it needs to give the Buyer the opportunity to test the data quality himself.

\noindent\textbf{Correctness:} Naturally one expects that the results that an algorithm produces are correct.
This needs to hold for this protocol as well.
Otherwise they are of no use and the Buyer cannot rely on the results when making his decision.
Since the graphs themselves are imperfect and cryptography is involved, insignificantly small and extremely unlikely errors are tolerable.

\noindent\textbf{Seller privacy:} In many cases the knowledge graph is of great value to the Seller.
Naturally he is not willing to give up any knowledge for free.
Seller privacy means that the Buyer only learns the information which the two parties agree upon.

\noindent\textbf{Buyer privacy:} Similarly to the Seller privacy, the Buyer does not want to reveal any information about his knowledge graph to the Seller.
Additionally, he needs to be sure that the Seller did not influence the outcome of the protocol.
If the Seller was able to influence the results, the Buyer would potentially be tricked into paying too much for the knowledge graph.

\noindent\textbf{Efficiency:} The size of real-world knowledge graphs range from very small to huge ones with billions of statements.
To be useful the protocol needs to be efficient and scale well with the size of the knowledge graphs it is run on.
However, because privacy is paramount and cryptography often is costly in terms of computation the protocol is allowed to take a while for huge graphs.

\section{Related Work}
\label{sec:relatedWork}
\todo[inline]{ 3/4 page}

Knowledge graphs are generally used to represent information about the world. 
The name was originally coined by Google to give a meaning to search terms instead of only matching key words~\cite{singhal2012introducing}.
In this work, we focus on RDF graphs~\cite{lassila1999resource}.
These graphs have nodes representing resources and directed, labeled edges indicating the relation between them.
These graphs can be serialized as a set of triples (see, e.g.,~\cite{Seaborne:14:RN}) called statements, where each triple (s, p, o) indicates one edge in the graph, by specifying the subject s (head), predicate p (label on the edge), and object o (tail).
We simplify RDF in the sense that we do not explicitly support named graphs nor blank nodes.


Knowledge graphs contain a certain amount of information. 
Measuring this information, and specifically how much information is gained by connecting two knowledge graphs together is a major aspect of this work.
Here, we were strongly influenced by the work of Sarasua et al.~\cite{sarasua2017methods}, who investigate the value of links between different data sets of a knowledge graph.
To do so, statistics (based on~\cite{auer2012lodstats}), such as the net amount of links, the average number of incoming and outgoing links per data set and number of entities linked by a source, are calculated.
In addition, other metrics based on Shannon entropy are computed for the knowledge graph with and without links between its data sets, where the difference between them serves as a metric for the information gained by these links. 
To the best of our knowledge these measurements, including the entropies, have not yet been investigated in a private setting.
Various works have looked into data quality in knowledge graphs, but the specifics are out of scope of this work.


In recent years there have been various famous examples of security breaches where huge amounts of sensitive data have been leaked.
This ranges from log-in data of websites to credit card details.
In the current work, we do not want either party to gain more knowledge as expected trough the proposed protocol, and hence a careful analysis of leaked information is needed.
A common approach in literature is to count the number of bits of 'sensitive' information which has been leaked.
In~\cite{borders2009quantifying} the maximum number of leaked bits in outbound traffic is measured.
Similarly, Backes~\cite{backes2009automatic} counts the number of leaked bits where the data is within some (case specific) logical equivalence class of sensitive data.
Vavilis~\cite{vavilis2014data} weights the leaked data such that the amount is not the only factor in determining the severity of a leak.
However, this weighting depends on the data and needs to be done by manually at least for a part of the data.
A similar approach is chosen in~\cite{vavilis2016severity} who also includes the anonymity of the leaked data as a factor. 
In our case though, a metric is needed that is independent of the concrete data, but specific to the case of knowledge graphs.
We elaborate on this in \cref{sec:Leaks}.


A Bloom filter is used for fast and efficient membership tests. 
It is a bit array starting with all zeros. 
Bloom filters are trained by hashing elements of a set with multiple hash functions to positions in the Bloom filter and setting them to 1.
To test if an element belongs to the set it gets hashed by all functions. 
If in every resulting position the Bloom filter contains a 1, the element is a member of the set, otherwise it is not. 
Because of collisions in hash functions false positives are possible, however false negatives are not.
The first time Bloom filters were proposed in \cite{bloom1970space} in 1970 by Bloom.
\cite{fan2000summary} introduced counting Bloom filters which are not just an array of bits but small counters instead. 
These counters get increased every time an element is hashed to this position. 
In our case Bloom filters will mainly be used to calculate set intersections.
A private set intersection protocol using Bloom filters is shown in \cite{nojima2009cryptographically}.


For our privacy preserving protocol, we build on top of existing techniques, like oblivious transfer and private set intersection.
Oblivious transfer was introduced by Rabin~\cite{rabin1981exchange} as scheme that allows Bob to retrieve a secret from Alice with a chance of 50\%. 
Alice does not find out whether Bob was successful or not.
Other variations include a scheme where Bob receives exactly one of two secrets from Alice without her knowing which one, for which a protocol has been presented in~\cite{even1985randomized}. This has then been generalized to Bob receiving one out of $n$ secrets.
An efficient scheme for this problem was presented in~\cite{tzeng2002efficient}. This scheme can be run in parallel to obtain a scheme for receiving $k$ out of $n$ secrets.
A direct and efficient $k$ out of $n$ scheme is presented in~\cite{chu2005efficient}.
Such protocols give Alice the guarantee that Bob cannot obtain more than $k$ secrets while Bob can be sure that Alice cannot learn which secrets Bob chose.


When agencies of different countries want to find matches in each other's criminal databases, or doctors want to find other patients with similar symptoms, they want to find the commonalities between datasets without disclosing information.
Often this gets reduced to the computation of a private set intersection, i.e., the computation of the intersection between sets where neither party is allowed to learn the information the other party has.
Different approaches exist for this problem.
For example, Freedman~\cite{freedman2004efficient} proposed an approach based on polynomials (later improved in~\cite{dachman2009efficient})
\todo[inline]{MC To save space: leave all OPRF parts out}
\todo[inline]{MC This is what I understood:}
The approach we follow in this work is based on Bloom filters as suggested by~\cite{nojima2009cryptographically}. 
Bloom filters are data structures very fast for membership-tests. 
They do, however, have the drawback of possible false positives.


In this work, we use the version with blind signatures, first introduced by Chaum~\cite{chaum1983blind}.
They are an important building block for many cryptographic protocols and used in various voting and payment schemes.
The most important characteristic is that the message which is signed is obscured or blinded.
Hence, the signer does not know what the message is that he signs.
Such a signature can then be publicly verified against the unblinded message. 
We use the blind signature scheme for RSA~\cite{bellare2003one}.
\todo[inline]{MC I like having the RSA description here, but there is too little space. Find something else to cut?}

\todo[inline]{This section could use a short ending statement}

\section{Approach}
\todo[inline]{Protocol centric explanation. 3 pages}

\label{sec:approach}

This section gives an overview of each of the steps taken in the protocol and how they work.
The protocol offers the opportunity for a Buyer to learn about the usefulness of a knowledge graph offered by a Seller.
The whole protocol is divided into five steps.
The first step is the initial contact when the parties agree on parameter settings and which parts of the protocol they want to use. 
During this step also some simple statistics in the knowledge graphs can be shared (but not verified).
In the second step, the intersection of the graphs is determined.
If the parties still want to continue after this step, a set of metrics (based on \cite{sarasua2017methods}) is computed, while still maintaining the graphs secret.
The outcome of the measures should give the Buyer an indication on whether the graph is interesting, but to determine the quality, actual data needs to be transferred (which could involve monetary exchange).
This happens in step 4 in such a fashion that the Buyer only obtains a subset of the data while the neither party can control which part that actually is.
Finally, if the Buyer is convinced the data is interesting and of sufficient quality, the deal is closed in step 5 of the protocol, where also all previous steps can be verified.
Note that either party can end of the protocol at each stage. 
Potentially because it has learned that the graph of the other party is not interesting after all.

\subsection{Step 1: Initial Contact} 

The goal of the initial contact is to determine how the protocol is run.
The Buyer states his interest and asks the Seller to start the protocol.
If there is interest from both sides, the protocol starts and they negotiate which of the metrics offered by the protocol they are willing to compute.
Each metric reveals something different about the Seller's knowledge graph and has its own trade off between information learned and potential information leaked.
Hence, they can make a decision based on the results of the information leak analysis section (\cref{sec:Leaks}) below.

At this stage the party (especially the Seller) might want to share specific information about his graph.
All of this information are aspects which can be computed from the graph without needed the other parties data, like simple statistics.
Examples include: the number of unique statements, resources, subject, predicates, or objects, as well as, in and out degrees, and so on (in our implementation we included 33 of these simple stats).
A further clarification can be given on the shared vocabularies used by the Seller.
Finally, the Seller might consider specific information too sensitive, or expensive, to share by chance (in step 4).
In that case they can agree that information can be removed or anonymized in an agreed upon fashion.
Note that none of the information shared can be verified at this stage; this is only possible at the very last step of the protocol.

\subsection{Step 2: Intersection} 
\label{sec:approach:intersection}

In this phase the Seller learns the intersection (i.e., the set of statments $i$ where $i \subseteq KG_S \land i \subseteq KG_B$) between the two knowledge graphs.
The goal is the evaluation of a private set intersection (PSI) scheme to reveal no more than the intersection to the Buyer.
To calculate the PSI an approach using a Bloom filter and blind signatures is chosen as proposed by Nojima~\cite{nojima2009cryptographically}.

First, the Seller computes a blind signature for each of his statements.
As he is now signing his own data, he can see it and just signs it.
Then, he computes a cryptographic hash of the concatenation of that signature with the original statement.
The outcomes get added to the Bloom filter.
When the buyer obtains this filter later, the Buyer can test whether statements he knows are in the Seller's graph (i.e., determine the intersection) by following the same steps.
However, the Buyer needs the Seller to sign the Buyer's statements, which are not visible to the Seller because of the blind signature approach.

The blind signatures prevent the Buyer from testing an infinite number of guesses against the Bloom filter, as all of them have to be signed off by the Seller, who will only sign a limited number (as agreed in step 1).
The secure hash-function further prevents the Buyer from reverse engineering statements from the Bloom filter.
\todo[inline]{to discuss later MC Isn't that already guaranteed by the previous signature step? Would a normal hash function be sufficient?}
The Seller sends the Bloom filter representation of his knowledge graph to the Buyer only after he has singed all blind signatures, he could also insist on doind the signing before sharing statisitcs.
This is to prevent cases where the Buyer would attempt to make informed guesses after he has learned information about the Seller's graph.

One issue remains, however. 
The Bloom filter might returns true for the membership test for statements not in the Seller's graph (i.e., yield false positives).
Therefore, it is important to choose a good filter size and number of hash functions when creating the Bloom filter to keep the false positive rate low.
If both parties deem it necessary, the false positive rate could also be kept high on purpose to give an idea of the intersection while still maintaining uncertainty for every positively verified statement.
This can either be achieved by choosing a smaller Bloom filter, more hash-functions or by randomly adding ones to the Bloom filter.

\subsection{Step 3: Entropy metrics}
\label{sec:approach:entropies}
\todo[inline]{MC Check: I try to call these consistently 'entropy metric' and not just entropy of the graph.}
\todo[inline]{The publication year on the Shannon entropy reference might be wrong. Check that one references are fixed}
After determining the intersection in the previous step, the goal of this step is to calculate metrics based on Shannon entropy \cite{shannon2001mathematical} to estimate the amount of information which can be gained from a merger of the two knowledge graphs.
These metrics were proposed in the work of Sarasua et al.~\cite{sarasua2017methods}.
The core idea is to contrast the entropy of a multi-set, derived from the graphs from multiple sources (i.e., the merged graph), with the entropy of a similar set derived from just one source.
One example of such metric is called the description of subjects (Desc.) of the graph.
In that case the multi-set consists of every predicate-object combination in the knowledge graph used to describe a statement.
\todo[inline]{MC If space and time left, add some more metrics.}
We implemented an analyzed this entropy metric computation for 9 different multi-set choices; the selection would in practice be determined in step one.
As we will discuss below, the calculation of the intersection is necessary for the correct calculation of the entropy metrics and therefore cannot be foregone unless one chooses to also not calculate any entropy metric, or allows an error while computing them.

For our privacy preserving setting, computing the metric for the single graphs is straightforward as no data has to be shared.
However, computing this for the merged graph is not obvious as the data cannot be shared.
What we will establish in this step is that there is no need to share the actual data.
The point is that it is sufficient that the Buyer knows the counts of elements in the merged multi-set, without knowing about the elements themselves.

The approach used is very similar to the one of the previous step, except that we use a counting Bloom filter instead of a normal one.
A counting Bloom filter differs from a normal one in that it keeps the count of the entities added to it, instead of only remembering their existence.
So, for each entropy metric separately, a counting Bloom filter will be send from the Seller to the Buyer, containing the counts for the given hash outcomes.
The Buyer now uses the blind signatures as before to obtain the hashes.
Since the goal is to obtain this entropy metric for the merged graphs, it is at this stage important to compensate for the intersection learned in step 2, as otherwise elements in the multi-set will be double counted.
\todo[inline]{later: MC There is some way described in the thesis, but it seems to me like it would be better to first subtract the intersection from the Buyer's graph end then compute the blind signatures, etc. Instead of doing the compensation in the end.
Hence, I did not include either explanation here.
 }
If the Buyer adds his own data (excluding the intersection) to the Bloom filter as well, the needed element counts can be obtained directly from the Bloom filter (again noting that no actual element information is needed, nor can be retrieved).
The difference of the entropy obtained from the Bloom filter (i.e., the entropy for the merged multi-sets), and the entropy of the Buyer multi-set gives an indication of the gain from merging the graphs.

Calculating an entropy for the combined knowledge graphs in this way can lead to a small error.
This error can originate from the false positive probability of the Bloom filter used in step 2 and 3.
Due to this, too many of the Buyer's statements might be considered as part of the intersection, leading to an overadjustment.
A second source of error is the false positive probability of the counting Bloom filter.
This can lead to elements of the Buyer being matched with elements of the Seller that are not the same.
However, choosing small enough false positive probabilities by making the Bloom filters big enough minimizes this problem. 
Further possible corrections, not implemented in our work because we could choose the Bloom filters large enough, are discussed by Cochez~\cite{DBLP:journals/corr/abs-1802-06609}.

For privacy reasons (which are discussed in detail in \cref{sec:Leaks}) the blind signatures for the Buyer must be handed out all at once and before the Seller sends over any of his (counting) Bloom filters, including the one from step 2.
This means one would compute all blind signatures for all entropies and the intersection in the previous step already.

After both step 2 and 3 are completed, the Buyer evaluates their outcome and decides if he is still interested in the Seller's knowledge graph.
If the values from step 2 and 3 are promising both parties can proceed with step 4. Otherwise the Buyer realizes that he cannot profit enough from buying the Seller's knowledge and the protocol ends with no deal between the two of them. \newline

\subsection{Step 4: Data Quality} 
\label{sec:approach:quality}

From the last two steps, the Buyer got a rough idea how much he could benefit from a merger of the two knowledge graphs.
The next important step is to ensure that the data is of acceptable quality.
This is very important as an adverse Seller could have put low quality or even nonsense statements into his graph to increase the values obtained in the previous steps.
However, determining the quality of the data is highly case specific and hence can only be done by looking at actual data.
Here, we assume that it is possible to determine the quality of the overall graph by obtaining a subset of the statements in the graph.
We assume that a deterministic procedure exists to partition the statements into $n$ equally meaningful sets and that obtaining $k$ of them at random makes quality checking possible; the actual check is out of scope of the current work.
Equally meaningful sets should be similar in size and representative (if possible) of the knowledge graph. Ideally they are also similar in monetary worth.
For our later evaluation we used both random partitions and other strategies which increase the likelihood that all statements concerning a given resource are withing the same partition. 

After the Seller's statements are partitioned in $n$ sets, we use a \emph{buy $k$ out of $n$ secrets} ($k$ oo $n$) oblivious transfer (OT) protocol, such that the Buyer obtains $k$ randomly selected parts of the knowledge graph for a certain predetermined price, after which he can investigate the quality at his own discretion.
Important is that this procedure ensures that neither party is able to determine beforehand which statements will be shared, implying that neither the Seller nor the Buyer can opt to include data of their choosing into what is shared.
Tampering at this stage would still be possible (for example, attempting to only including good data in the partitions), but this will be found out during the verification in step 5.

If it turns out that the graph is not of sufficient quality or does not contain real information the protocol may end here.
Otherwise, it continues with the final step. 

\subsection{Step 5: Closing the Deal and Verification} 

\label{sec:approach:verfication}
At this point the Buyer is sure that he wants to buy the knowledge graph.
Both parties negotiate a price for it, where the Buyer can use the results of the previous steps to judge how much the graph is worth to him.
Obviously, this is not part of the protocol.
Assuming they reach an agreement, they sign a contract and neither party can back out anymore.
This contract should contain specific clauses on what happens if it is not possible to verify all steps of the protocol, i.e., what happens if the Seller did not follow the protocol as expected; it is also possible, but not required to involve a third party at this stage of the protocol to perform the validation.
After this the Seller sends over his knowledge graph in plain-text and all private information he may have used for encryption, random number generation, etc.
during the protocol.
Based on this the Buyer can rerun the steps 2-4 on his own to make sure that the Seller has not cheated.
Naturally this only serves to detect a malicious Seller.
A curious Seller will not be exposed by this.

The procedure for verifying the computations is rather straightforward as it can exactly repeat the same steps as the protocol took.
The reason is that if all data is available, one can rerun the complete protocol without a second party.
However, as the privacy aspect is no longer important, some specific steps like encryptions can be skipped, reducing the computational burden during verification. 
For example, if one accepts to ignore small deviations of values, the computations using Bloom filters can be executed using normal sets and multi-sets, in that case there is not even a need to obtain the private key, used for the blind signatures, from the Seller.
Another example is the verification of the oblivious transfer, which can be sped up if the Seller sends all keys used in the process to the Buyer.

This step completes the protocol.\newline

\subsection{Implementation}

\label{sec:Implementation}

As a proof of concept and in order to evaluate the protocol, we implemented all steps in Java.
There is a program entry for both the Seller and the Buyer.
Even though most data is already enrypted in some form (e.g., the Bloom filters), we use TLS~\cite{dierks2008transport} to secure the communication, so that we do not have to perform a specific analysis for external eavesdroppers.

The applications read in their RDF graphs after which the protocol starts.
In interactive mode, the application asks after each major protocol step whether the user wants to continue.
Besides, the user has the possibility to select specifically which entropy metrics to compute.

For the computations of blind signatures we used RSA, as described in \cref{sec:relatedWork}.
The Seller needs to sign a lot statements, so we parallelized this step to use all available cores.
Because we know the number of statements which will be added to the Bloom filter, we can choose its size such that false positives rarely occur.

During the data quality step, several partitioning methods can be chosen.
\todo[inline]{MC Is the following correct?}
For our later evaluation we defaulted to what we called balanced DBSCAN which performs a graph clustering over the nodes, while attempting to keep the sets in the partitioning roughly of the same size.
The specifics of this are not important for the protocol itself.
We base the oblivious transfer implementation on the work of Chu and Tzen~\cite{chu2005efficient} but make use of RSA instead of Elgamal.
We do not directly feed the graph partitions into the oblivious transfer messages.
Instead, the partitions are encrypted (using cipher block chaining -- CBC ~\cite{dworkin2001recommendation} ) and the encryption keys are used in the messages.
One benefit of this is that the Buyer cannot choose to take partitions which look large, hoping to get more data.

At the end of the protocol, the Sender sends over the knowledge graph and all data needed for the verification, which is then performed by the Buyer.


\section{Protocol Analysis}
\todo[inline]{Describe properties, from section 5 and 6 in thesis. 2 pages}

\label{sec:Leaks}

The goal of this protocol is to evaluate the information gain that could be achieved when merging two knowledge graphs.
It is crucial to keep as much information private, i.e., until there is an agreement on sharing parts of the graph or selling it there should be as little information shared as possible.
This means that the Buyer learns only the agreed upon information about the Seller's knowledge graph.
At the same time the Seller should not learn anything about the Buyer's knowledge graph.
This section analyzes the safety of the protocol and which information gets leaked about either party.

In our analysis, we take the perspective of different types of adversaries. 
A non-participating adversary is not part of the protocol, but tries to learn something about the participants by attacking or influencing the protocol and its result. 
We neglect further analysis of this type as they would either have to find a weakness in either of the active parties' systems, which is not part of the protocol, or eavesdrop/attack the communication, which is part of the protocol but can be countered by classical cryptography which is not the focus of the paper. 

Hence, we only look into participating adversaries (i.e., the Seller and Buyer), as follows. 

\begin{itemize}[leftmargin=*]
	\item We will call either of the participating parties \textbf{Fair} if they are neither semi-honest/curious nor malicious.
	This means they have no bad intentions and stick to the protocol.
	\item A semi-honest \textbf{Curious} adversary is one of the participants of the protocol, and follows it strictly. 
	However, he is curious, i.e., tries to get as much information as possible, for example by combining information obtained from different steps.
	\item A \textbf{Malicious} adversary also takes part in the protocol, but aggressively attempts to gain information. This could be by faking input, skipping steps, or ending the protocol early. 
	\todo[inline]{MC I added the following sentence here. I am not sure this is true here... Remove if not confirmed.} 
	He does not care about being detected. 
	He is also curious.
\end{itemize}

Because only two parties participate in the protocol and it runs without a trusted third party, collusion of any kind is impossible.
A participating party does not gain any power by colluding with a non-participating adversary as it could behave like one itself.

\subsection{Information Leak Metrics}

To quantify how much information can (potentially) be leaked, we define the following metrics.

\noindent\textbf{ILStatements} is the number of parts of statements about which information is learned. Learning a complete statement increases the number with 3 (one for each of subject, predicate, and object).
	
\noindent\textbf{ILResources, ILSubjects, ILPredicates, and ILObjects} count the number of resources, subjects, predicates, and objects about which information is leaked, respectively.
These metrics are increased if it is discovered that the item is or is not contained in the other parties knowledge graph.
They are also increased if is found out how often the item occurs.

\noindent\textbf{ILStructural} This metric is a number indicating the number of pieces of structural information about the knowledge graph learned.
Such information is the size of the knowledge graph (number of statements), or the number of different resources, subjects, predicates, objects, or literals or the average number of in- and out-going links per node.
	
\noindent\textbf{ILAmount} This metric is the total information shared.

Note that the leaking of information can happen in either direction.
Hence, we subscript these metrics with $B\rightarrow S$ (e.g., $ILAmount_{B\rightarrow S}$) to indicate information leaked about the Buyer's knowledge graph which the Seller gains, and vice versa $S\rightarrow B$.
Further, we will write $X_{S\rightarrow B}$ or $X_{B\rightarrow S}$ to indicate all leak metrics in the specified direction.

Whether a metric is important, is case specific.
For example, if the parties have shared their ontologies beforehand, it might not matter much that $ILPredicates$ or $ILStructural$ is high.
Further, as discussed above, there could be cases where these numbers are not significant.
For example when the value of the graph is contained in a small subset of the statements.

In our analysis, each increase of a metric also increases the respective $ILAmount$ metric. 
We will not mention that explicitly.
Also, in some odd corner cases, like when the Seller would reveal that it has only one statement, the Buyer also learns that there is only one subject, predicate and object.
Similarly, if it is learned there is only one subject, and the number of statements, one knows its frequency.
We do ignore these corner cases as they are only of theoretical interest.

\subsection{Step 1: Initial Contact}

In the first step, it is immediately clear what is leaked, namely all intentionally shared statistics.
Each of these will add 1 to the $ILStructural_{S\rightarrow B}$ metric.
In our implementation we have altogether 33 of these statistics, so we continue assuming this step.
Note that even if it is decided that not all metrics are computed and shared, a non-fair Buyer might be able to compute the metric from the result of other ones.
Neither the Buyer, nor his knowledge graph are involved in the process and hence all his metrics stay zero.

The discussion about which parts of the protocol are to be run might give insight into which type of information is considered important by either party, but this is not part of the protocol itself and depends on the user's negotiations.

\subsection{Step 2: Intersection}

In this step information about the knowledge graphs is learned.
This step can be broken down into sub-steps, the first of which is the Seller filling the Bloom filter, which only involves the Seller, so no information is leaked.

In the second sub-step the Buyer obtains blind signatures for his statements, this does not involve the Seller's knowledge graph.
So all $X_{S\rightarrow B}$ are zero.
However, the Seller knows how many signatures he gives, therefore a curious Seller learns how many statements the Buyers knowledge graph contains.
Hence, $ILStructural_{B \rightarrow S} = 1$.\footnote{
	\todo[inline]{later: MC I added the footnote here. It would likely have been a good idea to add to the protocol from he start, but it might have effects I cannot investigate right now + not done in experiments.}
	
	If this is considered an issue, the Buyer can choose to get either more or fewer statements signed. 
	As these are blind signatures, the Seller cannot notice this anyway.
}
The remaining metrics remain zero, as the Seller does not learn what he signs~\cite{schroder2012security}.

In the final sub-step the Seller sends over the Bloom filter to the Buyer who tests it for the intersection.
The Buyer leaks nothing in this sub-step, but will gain new information.
A \textbf{fair} Buyer will learn the intersection and the subject, predicate and object for every statement in it.
Hence, assuming $n$ statements in the intersection\footnote{We consider the predicate as a potential additional resource here.}, $ILStatements_{S \rightarrow B} = 3n$, $ILResources_{S \rightarrow B} \leq 3n$, $ILSubjects_{S \rightarrow B} = ILPredicates_{S \rightarrow B} = ILObjects_{S \rightarrow B} \leq n$.
However, he learns nothing general about the graph, so $ILStructural = 0$. 
A \textbf{curious} Buyer can learn the size of the Seller's knowledge graph by counting the entries in the Bloom filter.
Also, if the intersection is large, he might deduce other structural information (like the number of resources or the average in- or outgoing links per node) with high confidence.
This results in $ILStructural_{S \rightarrow B} \geq 1$. The remaining metrics stay unchanged.
In the worst theoretical case the knowledge graph of the Seller is a subset of the knowledge graph of the Buyer and the whole graph is disclosed; meaning that there would also not have been any value for the Buyer in the first place.
Because the Bloom filter stores cryptographically secure hashes of the signatures the Buyer cannot reverse engineer the remaining statements from the Bloom filter.

Finally, the Buyer can also behave \textbf{maliciously}.
The only way this does let him gain new information is when the Buyer pretends to have certain statements, which he does not.
In that case, he can obtain blind signatures for them and in the end verify these against the Bloom filter.
After that, he would know that the Seller has these statements.
There are several countermeasures which can be used and factors which reduce the severity of the leakage.

First, it is generally not easy to guess statements without having any information about the graph \footnote{Experiments on guessing statement with real graphs will be provided in supplementary material.
\todo[inline]{MC todo: add experiment to supplement}
}.
So, the Seller can protect himself by not sharing too much information in step 1.
Besides, if there are still easy easily guessable statements with high value they can be removed or blinded as agreed in step one.
Then, this intersection step happens early in the protocol.
Therefore, information gained later in the protocol on which guesses could be based cannot be used (within the same run of the protocol).

Second, the Buyer only has a limited number of guesses. 
For each guess he needs the Seller to sign the statement and he will obviously only do that for a limited amount.

Third, the Seller could choose to provide a Bloom filter with a high false positive rate.
Then, even if a statement is confirmed, the Buyer cannot be certain it is in the Seller's graph.
This will obviously also affect the correctness of the intersection.

Last, if the above are not sufficient as a protection, the Seller can choose to not run this whole step.
The price to pay is that in the next step there is an error correction which becomes impossible.


\subsection{Step 3: Entropy Metrics}

The analysis of leaks while performing an entropy metric step is very similar to the previous one.
However, the information leaked is dependent on the specific metric computed, and hence each metric has its own leaks.
For the example metric we described above (\emph{`Desc'} in \cref{sec:approach:entropies}), the information which the Buyer might learn from the Seller's graph is summarized in \cref{ILSEntropyDesc}. For the other direction, only one piece of structural information can be recovered by a curious or malicious adversary.\footnote{We did analyze all 9 metrics for both parties, but left the analyses out in the interest of space and conciseness. These will be provided as supplementary material.}

\begin{table}
	\centering
			\todo[inline]{MC I changed this from Seller to Buyer. Pretty certain this should this be Fair Buyer, Curious Buyer, etc..}
	\begin{tabular}{|l|c|c|c|}
		\hline
		\multicolumn{4}{|c|}{Seller side information leaks: Desc}\\ \hline
		& Fair Buyer & Curious Buyer & Malicious Buyer \\ \hline
		ILAmount & $2$ & $\leq 2i_s + e_b + 3$ & $\leq 2i_s + e_b + 3$ \\ \hline
		ILStatements & $0$ & $\leq 2i_s$ & $\leq 2i_s$ \\ \hline
		ILResources & $0$ & $\leq \min(2e_s, 2e_b)$ & $\leq \min(2e_s, 2e_b)$ \\ \hline
		ILSubjects & $0$ & $0$ & $0$ \\ \hline
		ILPredicates & $0$ & $\leq \min(e_s, e_b)$ & $\leq \min(e_s, e_b)$ \\ \hline
		ILObjects & $0$ & $\leq \min(e_s, e_b)$ & $\leq \min(e_s, e_b)$ \\ \hline
		ILStructural & $2$ & $3$ & $3$ \\
		\hline
	\end{tabular}
	\caption{Seller side information leak during Desc entropy calculation.
	The cardinalities of the multi-sets are denoted with $i_s$ and $i_b$.
	$e_s$ and $e_b$ are the number of unique elements in the Seller's respectively Buyer's multiset. 
}
	\label{ILSEntropyDesc}%
\end{table}

\todo[inline]{later: MC Question: does the Buyer know which signed blind signature belongs to which original input? It seems like he does not need to know.
If he does not, this prevents certain attacks which try to guess elements.
This was further discussed on whatsapp:

The seller signs each element in his multiset and hashes them. He will send the Buyer a Bloom filter with this data.
The Buyer send encrypted elements to the Seller, together with the counts. The Seller does the blind signing. Then, he sends the Buyer the signed element and count pairs in a random order.
The Buyer gets the Sellers counting Bloom filter and can construct his own Bloom filter when removing the blinding from what he received from the Seller.
Now, the Buyer can add up the counts.

So, either the Seller learns the counts (but not the connection to elements) of the buyer OR the buyer can get to know the counts (and connection to elements) of the Seller.

This could work: if the Seller hashes the outcome of the blind signing. Would the Buyer then recover the encrypted thing from the Blinded thing.    unblind(hash(sign(blind(indexI)))) ?= hash (sign(indexI))

There is an issue with RSA, so this is currently not possible.

If I understood you right you might have the problem that rsa is only homomorphic if the Modulo stays the same. If you use two different rsa for the signing and then one for the "hashing" they would 2 different modulos. Otherwise the seller knows the primefaktors of the modulo (he is using for the signing) which would allow him to determine the second private key

If you have two rsa keys one with modulo N (First) and one with modulo M(second). Encrypt message x with the first one then with the second then decrypt with the first is not the same as only encrypting with the second one.
If both Modulo are the same then their prime divisors are the same. Then the seller would know them because He needs to know them to sign. If he knows them He can break the second rsa meaning he can build the private key

There might be some homomorphic hash function which could do this. 

}

\subsection{Step 4: Data Quality}

The only sub-step of the data quality step were there is interaction between the parties is when the Buyer obtains a subset of the partitioning using oblivious transfer.
Hence, at no other step information is leaked.
Further, because the whole oblivious transfer step is independent of the Buyer's knowledge graph there cannot be any Buyer side information leaks in this sub-step either.

The Buyer gets complete knowledge of those parts which are exposed by the transfer, but to nothing more~\cite{chu2005efficient}.\footnote{A detailed analysis is part of the supplementary material.}
Depending on the knowledge graph and the partitioning strategy it is possible for a curious Buyer to learn more structural information, by making the assumption that statistics for the parts obtained would also be valid for the complete graph.
A malicious one cannot do more than a curious one.
As these speculations are very unpredictable, case specific, and can be partially prevented by choosing a suitable partitioning strategy, we will not analyze them further.

\subsection{Step 5: Closing the Deal}

In this final step privacy is not an issue anymore for the Seller's graph.
During verification, the Seller is not involved, so there will be no leaks in that direction.

\todo[inline]{MC I left the totaling section and summary out for now. It builds on several of the metrics which have been left out as well.}

\section{Evaluation}
\label{sec:Evaluation}

\todo[inline]{MC 2 pages}

The analysis in the previous section, was aimed towards finding out what information is leaked in which step to make it possible for the Seller and Buyer to make an informed decision about whether they want to perform these steps.
Now, in this section, we show the results of experiments on realistic data sets, which indicate that the proposed protocol can be used in practice.
We run the complete protocol on knowledge graphs of different sizes and measure the run time, memory usage, and communication overhead.
Note, that we cannot show experimentally that the approach does not have security issues as this would require exhaustive experimentation, which is infeasible.

\noindent \textbf{Data Set} The data used for this evaluation is taken from knowledge graph ReDrugS used in~\cite{mccusker2017finding}. 
The knowledge graph contains information about Drug-Protein, Protein-Protein, and gene-disease interactions.
In this evaluation, we unify the thousands of named graphs into one default graph.
Next, we remove all statements containing blank nodes.
Then, a partition of this knowledge graph is created such that each part contain roughly the same number of statements and the internal sub-graph is connected.\footnote{This is done using a balanced DBSCAN-like clustering. See implementation for details.}
The graphs for the Seller and Buyer are created by merging randomly selected sets from the partition, trying to reach certain sizes.
We start with around 90,000 statements and increase in steps of 70,000 till we reach 400,000. 
Then we start increasing by 400,000 until we reach 2,800,000.
Note that the sizes of the graphs are not exactly rounded to these numbers.
Further, in our experiments we pair a Seller and Buyer knowledge graph of roughly the same size.
Four our reporting we use the average size of the Buyer and Seller graph.


\noindent \textbf{Run time environment} We run the experiments on two machines, one for each party.
Each of those machines has an Intel(R) Xeon(R) CPU E5-2640 v4 2.40 GHz with 20 cores.
The Java processes are limited to use 50 GB of RAM at most.
Both machines were connected to the same gigabit switch.

\subsection{Time}

\begin{figure*}
	\centering
	\includegraphics[width=\textwidth]{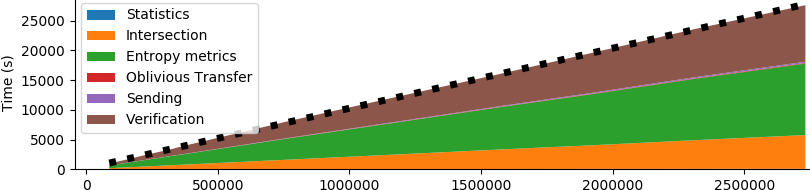}
	\caption{Run Time in function of the average number of statements}
	\label{chart:runtime-all}
\end{figure*}

The outcome of the timing measurements can be found in \cref{chart:runtime-all}.
It illustrates the time needed for each step of the protocol; but also includes `Sending', i.e., the time used to send over the knowledge graph and used keys from the Seller to the Buyer at the beginning of the 5th step.
What is shown is how the different steps of the protocol behave when dealing with differently sized graphs.
There are several observations:
\begin{enumerate}[leftmargin=*]
	\item The timing is dominated by the intersection, entropy metrics, and verification steps.
	This is pretty much as expected, as these are the parts which require a lot of encryption steps.
	Overall, the time is mainly consumed by computing blind signatures; it takes about 10 seconds to generate 10,000 blind signatures.
	
	\item The time needed grows linear in the size of the knowledge graph, as also indicated by the black dotted linear fit. 
	This is expected as most of the steps have an expected linear behavior.
	The only exception is the oblivious transfer, which has itself a step which in our implementation scales quadratically in the number of parts in the partition.
	However, even if we look at the OT step in isolation, we only see a linear behavior. Clearly, this quadratic step does not have a significant influence in practice.
	
	\item The time for verification of the complete protocol is about half of what is needed to run the protocol itself.
	
	\item For a graph containing close to 3 million statements, the time required is about 7.5 hours. 
	This time is definitely acceptable as this type of acquisitions would usually take several weeks to months anyway.
\end{enumerate}

\subsection{Memory}

\todo[inline]{MC For the figure, I made the assumption that the sizes in the excel file are in KB, rather than bytes (as indicated). Otherwise this would not make much sense.}

\begin{figure}
	
	\includegraphics[width=0.46\textwidth,right]{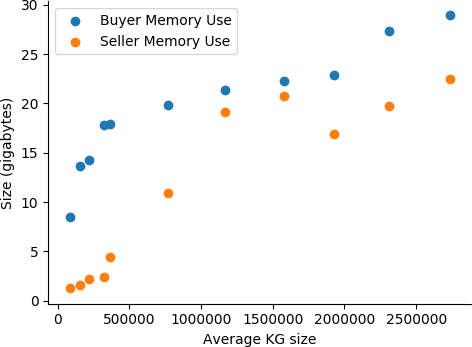}
	\caption{Amount of memory needed by the protocol in function of the average number of statements}
	\label{chart:memory_all}
\end{figure}

To determine the memory required by the protocol the peak amount of RAM used was tracked for different amounts of statements.
Note, that we do use a garbage collected JVM and hence, the memory usage can be rather spurious.
The measurements are summarized in \cref{chart:memory_all}.
The theoretical expectation is that the memory demand depends linearly on the number of statements.
We observe the following:

\begin{enumerate}[leftmargin=*]
	\item The Buyer always needs more memory than the Seller.
	\item The memory usage could be explained with a linear curve and behavior of the GC.
	\item The amount of memory needed is significant, meaning that our implementation cannot be ran on a laptop with low specifications at the moment.
	However, companies interested in using this protocol have typically access to large servers anyway or could rent these from a cloud service provider.
\end{enumerate}

More detailed testing showed that the intersection and entropy metrics steps take most memory.
In our implementation, for the Seller, there is also a significant amount of memory used in the partition step.
The cause is that we duplicate all data when creating the partition, while this would not be strictly necessary.
Finally, the verification also uses a lot of memory, but not as much as the private computations.

\subsection{Communication}

\begin{figure}
	\centering
	\includegraphics[width=0.5\textwidth]{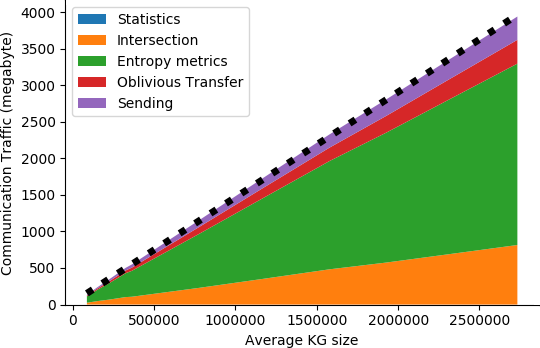}
	\caption{Amount of communication from the Seller to the Buyer in function of the average number of statements }
	\label{fig:communicationStoB}
\end{figure}

\begin{figure}
	\centering
	\includegraphics[width=0.5\textwidth]{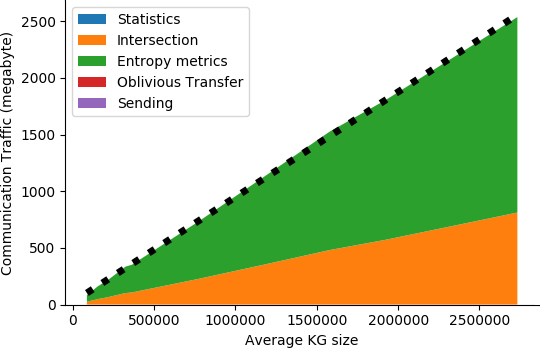}
	\caption{Amount of communication from the Buyer to the Seller in function of the average number of statements }
	\label{fig:communicationBtoS}
\end{figure}

The communication that was measured is the traffic that was generated by the protocol.
The measurements do not include overhead of the lower network stack (e.g., extra traffic caused by TLS or TCP).
We show the traffic going from Seller to Buyer in \cref{fig:communicationStoB} and in the other direction in \cref{fig:communicationBtoS}.
The verification step does not require any interaction between the two parties.
We observe the following:

\begin{enumerate}[leftmargin=*]
	\item It looks like each step the communication scales linearly with the number of statements, for both the Seller and the Buyer.
	The statistic step is, however, constant.
	\todo[inline]{MC I assume these numbers were computed on both experiments at the same time. Recompute for camera ready}
	The growth is roughly 0.47 KB per statement for the Buyer to Seller and 0.73 KB in the reverse direction.
	\item The trafic from the Buyer to the Seller is dominated by the Intersection and Entropy Metrics step. This is expected as other steps have either a very small (e.g., the keys for the oblivious transfer, linear in function of the number of parts) or constant overhead (e.g., statistics).
	\item Similarly, the traffic from Seller to Buyer is also dominated by these two steps. This is expected as all blinded statements to be signed by the Seller also have to be send back.
	For the Seller, also the oblivious transfer takes significant communication as he needs to send the complete encrypted graph to the Buyer.
	Finally, the Seller has to send the actual knowledge graph in the end, which is also a significant amount of data.
	
\end{enumerate}

Running experiments on the same data without any security measures reveals that the amount of communication grows 4.67 times bigger with our privacy preserving measures.


\section{Conclusion}
\todo[inline]{MC 1/2 page}

\label{sec:Conclusion}

Over the course of this paper a protocol was developed, which is capable of providing one party with information on how much it could learn form a second party's knowledge graph, without revealing the knowledge graphs of either party to each other or a trusted third party.
First, a set of statistics about the Seller's knowledge graph is computed and shared with the Buyer.
Then, the protocol makes use of blind signatures and Bloom filters to determine the intersection between the two knowledge graphs.
Similarly, counting Bloom filters are used to determine entropy metrics on the unification of both graphs, without either party needing access to more than its own knowledge graph.
To the best of our knowledge this is a new approach.
These two steps fulfill the requirement of information gain measurement.
In terms of correctness we have seen that there is a small error on the calculation of the intersection and the entropies.
Following this, the Buyer is given the opportunity to receive one or a few parts of the Seller's knowledge graph via oblivious transfer.
The technique works independent of the particular algorithm used for partitioning the knowledge graph.
In order to do so, strategies to partition knowledge graphs have been developed.
With these knowledge graph parts the Buyer can perform any data quality tests he deems necessary, thus satisfying the data quality requirement.
Finally, the Buyer has the chance to verify that the Seller did not cheat him. 

When analyzing the security of this protocol, we have seen that very little information, which is not supposed to be shared, gets leaked to the respectively other party.
This is not only true for curious but also malicious participants.
While not perfect this is pretty good in terms of the Seller and Buyer privacy requirements.
One problem is that while the protocol reveals little unwanted information, it gives a malicious Buyer the opportunity to verify guesses about the contents of the Seller's knowledge graph.
We have seen that by doing all blind signatures at the beginning, the Buyer is provided with no information by the protocol to base his guesses on. 
Besides, several further mitigation options were discussed.

This protocol has then been implemented in Java.
The evaluation of this implementation has then shown, that it scales linear in time, communication and presumably memory usage with the number of statements on the side of Seller and Buyer.
It has also shown, that blind signatures are responsible for most of the time used, even though they can be computed in parallel.
For big knowledge graphs the time and memory needed can grow quite large.
The linear scaling satisfies the requirement of efficiency. 

\subsection{Future Work}

In future work different areas of this protocol can be further investigated.
First of all, the implementation can be adapted to also handle knowledge graphs with blank nodes and containing named graphs.
Named graphs would open new possibilities for entropies and can be an alternative to the partitioning needed in the data quality step.
Another point of improvement is finding a way to reduce the computation time for the blind signatures.
It seems hard to substantially improve the time needed, though.
Possibly using a different private set intersection approach could bring a speedup to the intersection and the entropy step. 

For the entropy metrics calculation, there might be a way to reduce the amount of leaked information further.
The goal is to compute only a number, but also information about the Seller's knowledge graph is leaked.
The main reason is the Buyer knows (needs to know in our protocol) which elements in the multi-sets are in common to be able to match them with his own data.
Perhaps some technique involving homomorphic encryption could solve this problem, but this will require the development of a new blind signature protocol.

Partitioning the knowledge graphs for step 4 could also be an area for future work.
Efficiently partitioning a knowledge graph in such a way that each part is of the same or similar value is a very challenging task.

In terms of security the guessing can be a problem but seems nearly impossible to prevent.
Because of the blind signatures the Seller can never know if he is signing an actual statement or element of the Buyer or a guess.
Also, finding a way to automatically determine the number of signatures the Seller should be willing to hand out seems unrealistic. 

\newpage

\bibliographystyle{ACM-Reference-Format}
\bibliography{bibliography}

\end{document}